\documentstyle[12pt,epsfig]{article}
\newcommand{\ep}{\smallskip}

\newcommand{\be}{\begin{equation}}
\newcommand{\beo}{\begin{displaymath}}
\newcommand{\eeo}{\end{displaymath}}
\newcommand{\ee}{\end{equation}}
\newcommand{\bes}{\begin{eqnarray}}
\newcommand{\ees}{\end{eqnarray}}
\newcommand{\beso}{\begin{eqnarray*}}
\newcommand{\eeso}{\end{eqnarray*}}
\newcommand{\bma}{\left( \begin {array}}
\newcommand{\ema}{\end {array} \right)}
\newcommand{\pslash}{\kern 0.2 em p\kern -0.45em /}
\newcommand{\dslash}{\kern 0.2 em \delta\kern -0.45em /}
\newcommand{\sla}[1]{\kern 0.2 em #1\kern -0.45em /}

\begin{document}  

\begin{center}
{\bf Thermodynamics of the Casimir effect}
\end{center}
\vspace*{1cm}
\begin{center}
H.~Mitter and D.~Robaschik
\\
Institut f\"ur Theoretische Physik, Karl-Franzens-Universit\"at Graz, \\
Universit\"atsplatz 5, A-8010 Graz, Austria
\vspace*{1cm}
\end{center}

A complete thermodynamic treatment of the Casimir effect is presented. Explicit
expressions for the free and the internal energy, the entropy and the pressure
are discussed. As an example we consider the Casimir effect with different
temperatures between the plates ($T$) resp. outside of them ($T'$). For $T'<T$ the 
pressure of heat radiation can eventually compensate the Casimir force and the 
total pressure can vanish. We consider both an isothermal and an adiabatic 
treatment of the interior region. The equilibrium point (vanishing pressure) 
turns out instable in the isothermal case. In the adiabatic situation we have 
both an instable and a stable equilibrium point, if $T'/T$ is sufficiently 
small. Quantitative aspects are briefly discussed.

\medskip\noindent
PACS: 05.90.+m, 11.10.Wx, 61.16.Ch

\newpage
\begin{flushleft}
{\bf 1. Introduction}
\end{flushleft}

The Casimir effect \cite{CA} is one of the fundamental effects of Quantum Field
Theory. It tests the relevance of the zero point energy: two parallel, large
conducting plates in a distance $a$ change the vacuum energy of Quantum Electrodynamics
in such a way, that a net attractive force between the plates results. 
Qualitatively, this situation does not change, if the system is at finite temperature.
The quantitative corrections for finite temperature \cite{FM}, finite conductivity
\cite{SRM}
and the necessary changes for related problems (e.g.for different geometries) 
\cite{MO} are known,
and the Casimir force has been experimentally established \cite{EX1,EX2,EX3}. 
The thermodynamics
for the original setup has been investigated previously \cite{FM,BM,SRW}. Here
we shall give a somewhat more complete treatment, in which we consider the region
between the plates and outside of them separately. This allows for a situation, in
which the two regions have different temperatures (outside $T'$, inside $T$). If we take 
$T'<T$, the external pressure is reduced in comparison with the standard situation
$(T'=T)$. Therefore we expect the existence of a certain distance $a_0$, at which the
Casimir attraction is compensated by the net radiation pressure. 
We shall investigate this (mechanical) equilibrium point and its stability 
both for an isothermal and an adiabatic treatment of the interior region.
\begin{flushleft}
{\bf 2. Thermodynamic functions}
\end{flushleft}

In this section we shall collect and discuss briefly formulae for thermodynamic
functions for the original Casimir setup. A major part of the results has been
found before \cite{BM,SRW} by various methods. In order to make this paper
self-contained, we shall indicate briefly a standard procedure for the derivation
and list the results. In addition, we shall give numerical details on the functions
involved. We consider two parallel, perfectly conducting square plates (side $L$,
distance $a$, $L>a$), embedded in a large cube (side $L$) with one of the plates
at face and periodic boundary conditions. We shall consider contributions from the
volume $L^2a$ between the plates (suffix int) resp. $L^2(L-a)$ outside of them (suffix
ext) separately. In terms of the partition function $Z$ the free energy of a photon
gas at temperature $T$ reads
\be
F = - \frac{1}{\beta} \ln Z = \sum (G_0 (k) + G_{\beta} (k))
\ee
with
\be
G_0 (k) = \frac{\hbar c}{2} k,\ \ \  \ G_{\beta} (k) = \frac{1}{\beta} \ln \left( 1 -
\exp \left( - \beta \hbar ck \right) \right).
\ee
Here $\beta = 1/k_B T$ with Boltzmann's constant $k_B$ and $\hbar c k = \hbar c
\sqrt{{\bf k}^2}$ is the energy of a photon with wave number ${\bf k}$. The sum
extends on all accessible wave numbers and polarisation states. In order to keep
track of ultraviolet divergencies, we shall regularize the contribution at
temperature zero replacing $k \rightarrow k \exp \left( - \lambda k \right)$ 
in $G_0$ and considering $\lambda \rightarrow 0$ at the end. We assume $L$
fixed, but so large, that sums on photon wave numbers can be replaced by corresponding
integrals. Then the contributions to the free energy per unit area $\phi = F/L^2$ read.
\be
\phi_{\rm ext} = \frac{(L-a) \pi}{a^3} \int_0^{\infty} t^2 G \left( \frac{\pi}{a} t
\right) dt, 
\ee
\be
\phi_{\rm int} = \frac{\pi}{a^2} \left( \frac{1}{2} \int_0^{\infty} + \sum_{n=1}^{\infty}
\int_n^{\infty} \right) t G \left( \frac{\pi}{a} t \right) dt.
\ee
The integrals can be done analytically. In the results to be listed below we have
expanded the result for $G_0$ in powers of $\lambda$ and omitted all terms, which
vanish for $\lambda \rightarrow 0$. For $G_{\beta}$ it turned out convenient to
use the polylogarithm
${\cal L}_n (z) = \sum_{m=1}^{\infty} z^m/{m^n}$, 
which allows even for a calculation of the indefinite integrals and has simple
properties (cf. Appendix). Further thermodynamic functions can be calculated 
from $\phi$ by standard formulae. We shall give results for the entropy
\be
\sigma = S/k_B L^2 = \beta^2 \frac{\partial}{\partial \beta} \phi,
\ee
the pressure
\be
P= - \left(\frac{d\phi}{da}\right)_T
\ee
and the internal energy
\be
\epsilon = \frac{E}{L^2} = \frac{1}{L^2} (F+TS) = \phi + \frac{1}{\beta} \sigma.
\ee 
The results are listed in Table 1 in terms of a finite function $g(v)$ to be
specified below. The variable $v$ is the dimensionless quantity 
(cf. \cite{BM,SRW,ROM})
\be
v = \frac{a}{\pi \hbar c \beta} = \frac{a k_B}{\pi \hbar c} T.
\ee
\begin{tabular}{|c|l|l|} \hline
\multicolumn{3}{|c|}{Table 1} \\ \hline
function & ext  contribution & int contribution \\ \hline
$\phi$ & $\left( L-a \right) \hbar c \left[ \frac{3}{\pi^2 \lambda^4} - \frac{\pi^6}{45}
( \frac{v}{a} )^4 \right]$ & $\hbar c \left[ ( \frac{3a}{\pi^2 \lambda^4} +
\frac{\pi^2}{a^3} ( - \frac{1}{720} + g(v) )\right].$ \\ \hline
$\sigma$ & $\left( L-a \right) \frac{4 \pi^5}{45} \left( \frac{v}{a} \right)^3$ &
$- \frac{\pi}{a^2} \ g' (v)$ \\ \hline
$P$ & $\hbar c \left[ \frac{3}{\pi^2 \lambda^4} - \frac{\pi^6}{45} ( \frac{v}{a}
)^4\right]$ & $\hbar c \left[ - \frac{3}{\pi^2 \lambda^4} + \frac{\pi^2}{a^4}
( - \frac{1}{240} + 3 g(v) - v g' (v) ) \right].$ \\ \hline
$\epsilon$ & $( L-a ) \hbar c \left[ \frac{3}{\pi^2 \lambda^4} + \frac{3Ê\pi^6}{45}
( \frac{v}{a} )^4 \right]$ & $\hbar c \left[ \frac{3a}{\pi^2 \lambda^4} + 
\frac{\pi^2}{a^3} \left( - \frac{1}{720} + g(v) - v g' (v) \right)\right].$ \\ \hline
\end{tabular}
\vspace*{0.5cm}

The ultraviolet behavior can be discussed without any details on $g$. Both contributions
to $\sigma$ are finite. The divergent contributions to the pressure cancel in the sum
$P_{\rm ext} + P_{\rm int}$ and in the difference $P_{\rm ext} (\beta') - P_{\rm ext}
(\beta)$, which we shall use in the next section. The total free energy $\phi_{\rm ext}
+ \phi_{\rm int}$ contains a divergent constant $\sim L/\lambda^4$, as well as the total
internal energy.

Another formula, which can be read off Table 1 without knowledge of $g$ is the 
relation
\be
3 \phi + \frac{1}{\beta} \sigma - aP = L \hbar c \left( \frac{9}{\pi^2 \lambda^4} +
\frac{\pi^6}{45} \left( \frac{v}{a} \right)^4 \right),
\ee
which is connected with the fact, that physically relevant contributions scale as $b(v)/
a^3$ with some function $b$.

The function $g(v)$ contains an infinite sum, which cannot be performed analytically.
Various forms can be found in the literature \cite{BM,SRW}. We shall prefer a
representation, which allows for a fast and accurate numerical computation. For this
purpose we use the functions (cf. Appendix)
\be
k(x) = \left(1-x\frac{d}{dx}\right) \sum_{n=1}^{\infty} \frac{1}{n^3(\exp (nx)-1)},
\ \ h(x)=xk'(x) 
\ee
For $x>0$ the sum converges rapidly, $k$ is positive and $h$ is negative. Both 
functions vanish exponentially for
large argument. Two representations of $g$ in terms of $k$ read 
\be
g(v) = - v^3\left( \frac{1}{2} \zeta (3) + k(1/v)\right),
\ee
\be
g(v) = \frac{1}{720} - \frac{\pi^4}{45} v^4 - \frac{v}{4\pi^2}\left( 
\frac{\zeta (3)}{2}+k(4\pi ^2 v)\right).
\ee
with $\zeta (3) ={\cal L}_3 (1) =1.2020569\cdots $. 
The two forms are related by Poisson's sum formula \cite{Ti}. Taken together, they
contain information on the behavior of thermodynamic functions under temperature
inversion $(T \to 1/T)$ \cite{BM,RT} (the quantities used in \cite{BM} are 
$\xi = \pi v, \ f(\xi) =\pi^6 v^4 /45 + \pi^2 g(v)$). 

In order to discuss the 
behavior of the thermodynamic functions quantitatively, we write
\be
\phi = \phi_L (T) + \frac{\pi^2 \hbar c}{a^3} f(v), \ \sigma = \sigma_L (T) +
\frac{\pi}{a^2} s(v), \ P= \frac{\pi^2 \hbar c}{a^4} p(v), \ 
\epsilon= e_L (T) + \frac{\pi^2 \hbar c}{a^3} e(v),
\ee
where ($\phi_L,\sigma_L,e_L$) refer to the extensive contributions $\sim L$ 
from the external cube. The remainders are 
connected by simple relations. As a consequence of (9) we have
\be
3f(v)+vs(v)-p(v)=0.
\ee
Equ. (7) amounts to
\be
e(v)=f(v)+vs(v).
\ee
Therefore only two of the four functions $(f,s,p,e)$ are linearly independent. 
Explicit forms in terms of $g$ can be read off Table 1. 
Expressions in terms of $k(x),h(x)$  
result by insertion of $g$. There are two equivalent forms (A,B) for 
every function corresponding to (11) resp. (12), which are listed in Table 2.

\begin{tabular}{|c|l|l|} \hline
\multicolumn{3}{|c|}{Table 2} \\ \hline
funct.&form (A)&form (B) \\ \hline
$f$&$-\frac{1}{720}-v^3\left( \frac{\zeta (3)}{2}-\frac{\pi ^4}{45}v+k(1/v)\right)$
&$-\frac{1}{4\pi ^2}v\left( \frac{\zeta (3)}{2}+k(4\pi ^2v)\right) $ \\ \hline
$s$&$v^2\left( \frac{3\zeta (3)}{2}-\frac{4\pi ^4}{45}v+3k(1/v)-h(1/v)\right)$
&$\frac{1}{4\pi ^2}\left( \frac{\zeta (3)}{2}+k(4\pi ^2v)+h(4\pi ^2v)\right)$
 \\ \hline
$p$&$-\frac{1}{240}-v^3\left( \frac{\pi ^4}{45}v+h(1/v)\right)$
&$-\frac{1}{4\pi ^2}v\left( \zeta (3)+2k(4\pi ^2v)-h(4\pi ^2v)\right)$
 \\ \hline
$e$&$-\frac{1}{720}+v^3\left( \zeta (3)-\frac{3\pi ^4}{45}v+2k(1/v)-h(1/v)\right) $
&$\frac{1}{4\pi ^2}vh(4\pi^2v)$ \\ \hline
\end{tabular}
\vspace*{0.5cm}

The leading terms for small values of $v$ (i.e. at low temperatures) are found 
from forms (A) neglecting $(k,h)$. This approximation can be used for $v<0.09$ 
with errors less than 1\%. The asymptotic behaviour at large values of $v$ 
(i.e. at high temperatures) is obtained from forms (B) neglecting $(k,h)$. The 
corresponding asymptotic approximations for $(f,s,p)$ can be used in a rather 
large domain. For $v=0.25$ the 
errors are $\leq $1\% and decrease rapidly with growing $v$. 
Form (A) shows, that the entropy vanishes for $T=0$, i.e. Nernst's law is 
fulfilled. On the other hand $T$ tends to a finite value $T(a=0)$ for $a\to 0$, 
which is determined by the entropy alone.  
Form (B) shows, that the entropy becomes constant in the 
high-temperature limit (Kirchhoff's theorem \cite{ROM}). It has to be noted, 
that there is only a rather narrow domain in $v$, in which the infinite sums 
$(k,h)$ play a role. In 
this region the low-temperature behavior of the 
Casimir force is gradually changed by temperature effects. This domain 
is realistic \cite{EX2,EX3}: for $T=18^0{\rm C}$ we have 
$v\approx 0.04a(\mu {\rm m})$. 
Using the expression (10), one can find numerical values in this domain rapidly 
and with high precision. Results are shown in the plots given in Fig.1 
(in which the broken lines correspond to the asymptotic approximations). 
\begin{figure}[hbt]
\begin{center}
\epsfbox{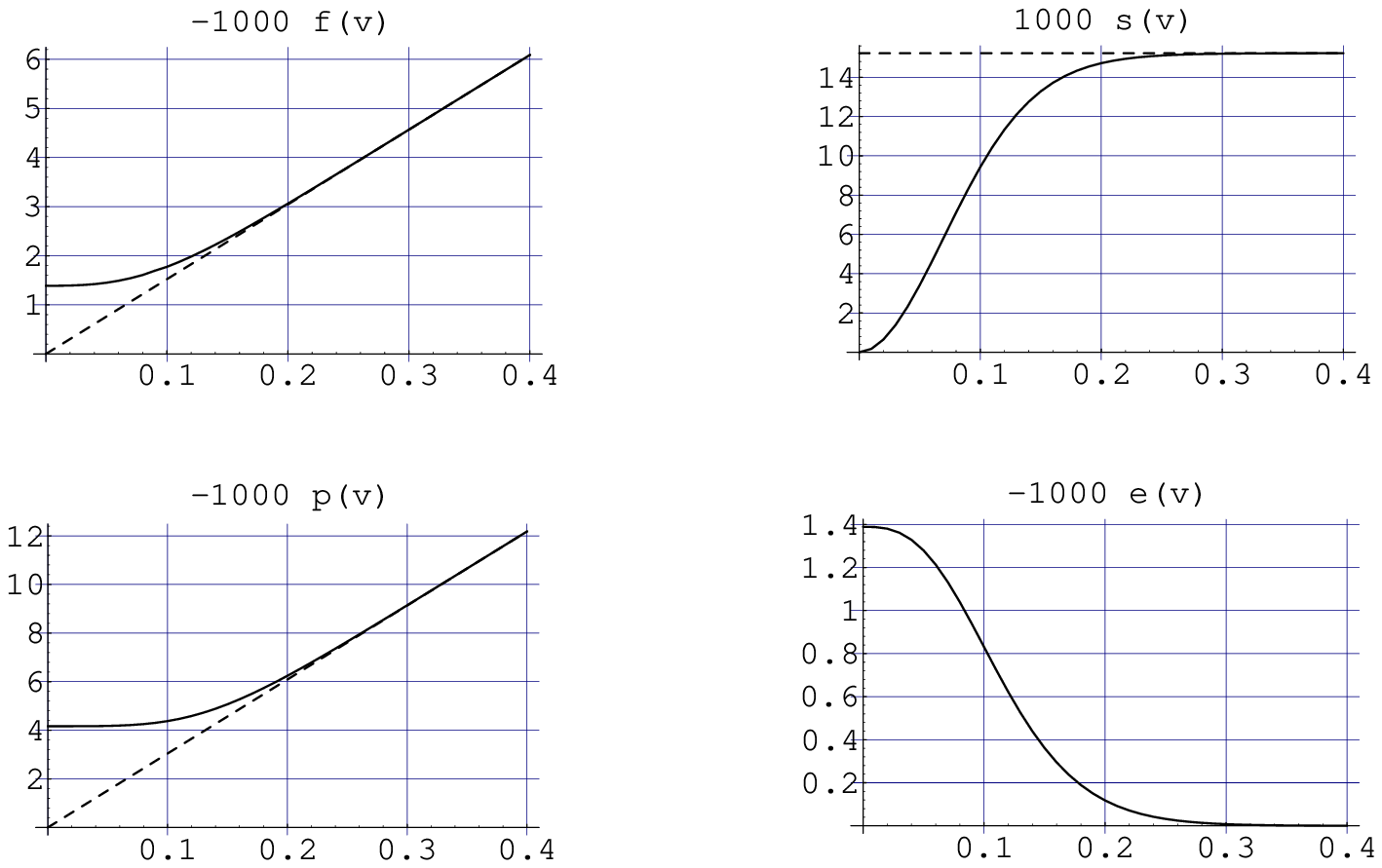}
\end{center}
\end{figure}

\begin{center}
{\small Fig.1 Thermodynamic functions. }
\end{center}

\begin{flushleft}
{\bf 3. Different Temperatures and Equilibrium}
\end{flushleft}
\vspace*{0.2cm}
As an application we shall now investigate a situation, in which the temperature 
$T$ between the plates may differ from the temperature $T'$ of the exterior region. 
We shall keep $T'$ resp. $v'=ak_BT'/\pi \hbar c$ fixed. For the interior region 
we consider two possibilities. In the first case also $T$ resp. $v$ is kept fixed 
(isothermal situation). In the second case we treat the internal region as a 
closed system, so that the entropy $\sigma _{int}$ remains constant (adiabatic 
situation) and $T$ resp. $v$ vary appropriately with $a$. In both cases we shall focus 
attention on the question, whether there is a distance $a_0$, at which the total 
pressure vanishes. If one of the two plates is movable, the distance $a=a_0 $ 
would constitute mechanical equilibrium between the Casimir attraction and 
the thermal radiation pressure (which repels the plates). We shall also consider 
the question, whether this equilibrium is stable.  

For an appropriate treatment we consider the quantity
\be
P(a,v,v')=P_{ext}(T')+P_{int}(T)=\frac{\pi ^6\hbar c}{45a^4}G(v,v')
\ee
with 
\be
G(v,v')=v^4+\frac{45}{\pi ^4}p(v)-v'^4
\ee
in terms of $p$ from section 2. We shall always use the form 
(B) from Table 2. The equilibrium condition
\be
P(a,v,v')|_{a=a_0}=0
\ee
amounts to
\be
G(v,v')|_{a=a_0}=0.
\ee
Solutions are only possible, if $v'<v(a)$, since $p$ is negative.
A Taylor expansion of $P$ near $a=a_0$ shows, that $dP/da$ at equilibrium 
must be negative for a stable situation. Differentiating (16) and using (19), 
one observes, that the stability condition amounts to
\be
\frac{dG(v,v')}{da}\Big|_{a=a_0}<0.
\ee
We shall always seek solutions considering a scaled form of (19)
\be
\kappa ^4=R(v),\ \ R(v)=\frac{1}{v_0^4}\left( v^4+\frac{45}{\pi ^4}p(v)\right),
\ \ \kappa =\frac{v'}{v_0}
\ee
with some scale $v_0$.

In the {\bf isothermal} case $v$ is proportional to $a$. If we take $v_0=v$, 
the ratio $\kappa =T'/T$ does not depend on $a$. From the plot for $p(v)$ 
given in Fig. 1 we can see, that $R(v)$ is a monotonous function 
with positive tangent. The function is positive above a certain value of $v$. 
For any given value of $0\leq \kappa \leq 1$ we have therefore one solution, which is 
unstable. Some numerical results are listed in Table 3. The third row gives 
the distances for $T=18^0 {\rm C}$ in $\mu {\rm m}$. The last row is obtained 
using the asymptotic approximation $p_{as}=-v\zeta (3)/4\pi ^2$, 
which turns out quite accurate.
\ep
\begin{center}
\begin{tabular}{|l|l|l|l|l|l|l|} \hline
\multicolumn{7}{|c|}{Table 3: isothermal equilibrium}\\ \hline
$\kappa $&0&0.2&0.4&0.6&0.8&0.95\\ \hline
$v(a)$&0.2419&0.2421&0.24340&0.2532&0.2879&0.4233\\ \hline
$a(18^0{\rm C})/\mu {\rm m}$&5.981&5.984&6.032&6.260&7.117&10.46\\ \hline
$v_{as}(a)$&0.2414&0.2415&0.2435&0.2528&0.2877&0.4233\\ \hline
\end{tabular}
\end{center}
\ep
In the {\bf adiabatic} case we have to evaluate (21) along an 
isentrope, which corresponds to a curve $\sigma_{int}(v)={\rm const}$. A convenient 
parametrisation is obtained using the constant
\be
t^2=\frac{2\sigma_{int}}{3\pi \zeta (3)}
\ee
instead of $\sigma_{int}$. An equation for the isentrope is obtained 
inserting the explicit expression for $\sigma_{int}(v)=-\pi g'(v)/a^2$ from 
section 2. In terms of $t$ and $s$ the equation reads
\be
a^2t^2=\frac{2}{3\zeta (3)}\left( s(v)+\frac{4\pi ^4}{45}v^3\right).
\ee
The leading term on the r.h.s. for small $v$ is $v^2$ (see Table 2, form (A)). 
Therefore the isentrope becomes an isotherme for small $v$ and $t$ can 
be identified as
\be
t= \frac{k_B}{\pi \hbar c}T(a=0).
\ee
\begin{figure}[hbt]
\begin{center}
\epsfbox{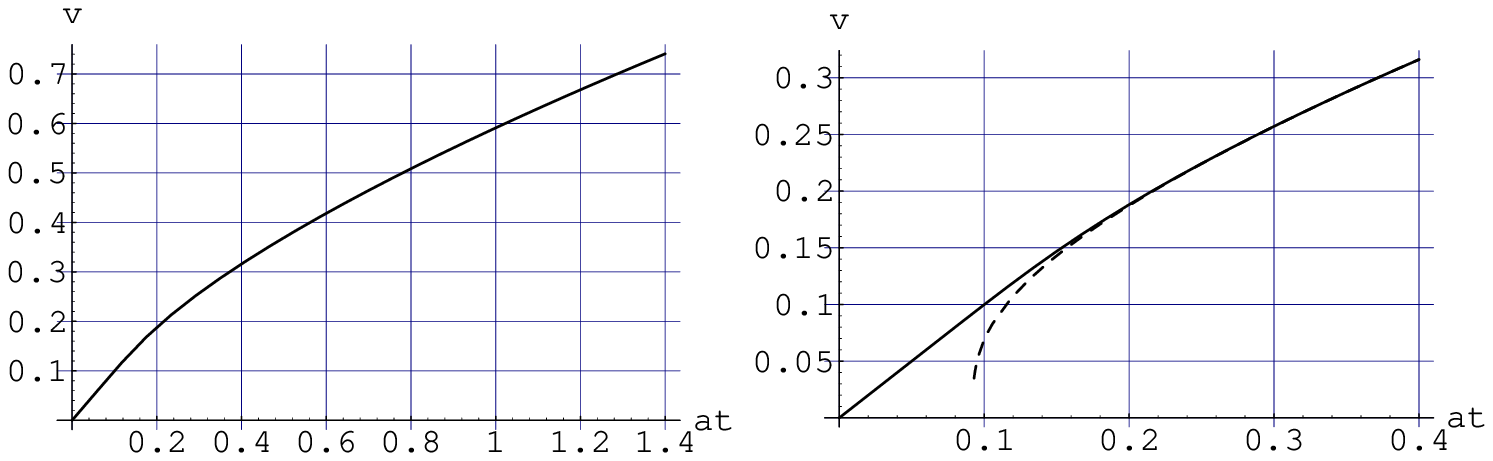}
\end{center}
\end{figure}

\begin{center}
{\small Fig.2 Isentrope. }
\end{center}

In order to obtain an explicit form for the isentrope we have to solve 
(23) for $v=v(at)$. This can be done numerically with high precision. 
We have used the form (B) from Table 2 for $s(v)$ to determine $v(at)$. 
The result is plotted in the left diagram of Fig.2.
An approximate form $v_{as}(at)$ is obtained neglecting $k+h$ 
in $s(v)$. The result is
\be
v_{as}(at)=\left( B(12\pi ^2a^2t^2-1)\right) ^{1/3},\ \ B=\frac{45\zeta (3)}{32\pi ^6}.
\ee
$v$ can be approximated by $v_{as}$ for $at\geq 0.25$ with an error $\leq 0.1$\%. 
With increasing $at$ the error decreases rapidly. The right diagram of Fig.2 
(in which the broken line corresponds to $v_{as}$) illustrates this fact.

If we take $v_0=at$ in (21), the ratio    $\kappa (0)=T'/T(a=0)$ does not depend on 
$a$ and can therefore be used as an input for (21). The function $R(v)$ shows, 
however, a different behavior than in the isothermal case. It vanishes for 
$at=0.2763$, assumes a maximum at $at=at_M=0.4391$ 
and falls off slowly for larger argument (see Fig.3).
\begin{figure}[hbt]
\begin{center}
\hspace{-0.6cm}
\epsfbox{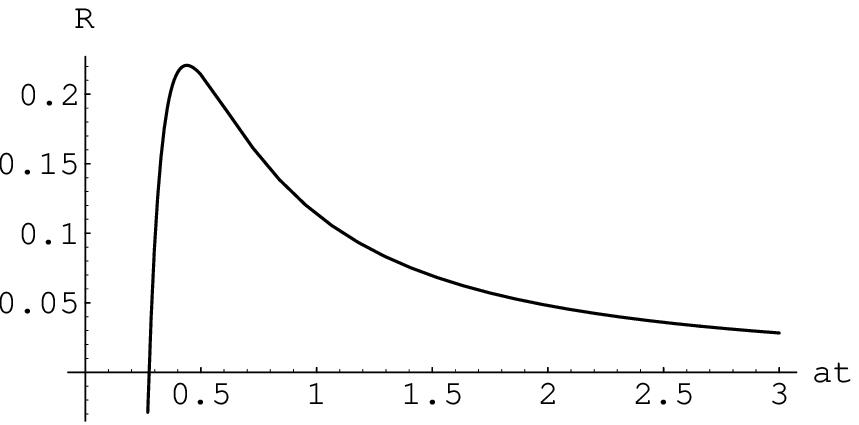}
\end{center}
\end{figure}

\begin{center}
{\small Fig.3 Adiabatic equilibrium function R.}
\end{center}

Therefore we obtain no solution, if 
$\kappa (0)$ is larger than $\kappa _M=(R(at_M))^{1/4}=0.68542$. For $\kappa <\kappa _M$ 
there are two solutions: one of them (corresponding to the lower value of 
$at$) is unstable, the other one is stable. The distance between the two 
solutions increases with decreasing $\kappa (0)$. Some numerical results are 
listed in Table 4. The entries for $a(18^0{\rm C})$ refer to $T(0) =18^0{\rm C}$.
\ep
\begin{center}
\begin{tabular}{|l|l|l|l|l|l|} \hline
\multicolumn{6}{|c|}{Table 4: adiabatic equilibrium}\\ \hline
 &$\kappa (0)$&0.2&0.4&0.6&0.65\\ \hline
unstab.&$at$&0.2767&0.2819&0.3148&0.3441\\ \hline
 &$a(18^0{\rm C})/\mu {\rm m}$&6.839&6.969&7.783&8.506\\ \hline
 &$\kappa (a)$&0.2285&0.4592&0.7093&0.7876\\ \hline
stab.&$at$&26.03&3.235&0.8917&0.6503\\ \hline
 &$a(18^0{\rm C})/\mu {\rm m}$&643.4&79.98&22.05&16.08\\ \hline
 &$\kappa (a)$&0.99998&0.9984&0.9778&0.9565\\ \hline
\end{tabular}
\end{center}

It is observed, that $\kappa (a)$ is close to 1 for the stable equilibrium 
points. This is due to the slow falloff of $R(v)$. Using the asymptotic 
approximation $p\to -v_{as}(at)\zeta (3)/4\pi ^2$, the data in Table 4 
are reproduced with rather small errors.
 
The existence of stable equilibrium points implies, that a movable plate 
may perform oscillations about these points. In order to approach the 
problem of an observation, we shall discuss briefly some quantitative details. 
For these, we have used the formula for the Casimir pressure 
\be
\frac{\pi ^2\hbar c}{240a^4}=1.3001\cdot 10^{-7} \frac{{\rm N}}{{\rm cm}^2}
\ee
Here $a$ is measured in microns ($\mu $m). We assume the movable plate 
situated at the distance $a_1$ from a stable equilibrium point $a$, where 
$a_1/a$ is small enough, so that a linear approximation is possible. 
For the data in Table 4 (stable points, $18^0$C) we obtain a restoring 
force/area of
\be
\frac{a_1}{a}(0.39\cdots 28)\frac{{\rm pN}}{{\rm cm}^2}.
\ee
The oscillation frequency is
\be
\nu = \frac{1}{\sqrt m}(0.12\cdots 6.7)\cdot 10^{-3}{\rm Hz}.
\ee
Here $m$ is the mass per area of the movable plate measured in g/${\rm cm}^2$. 
Thus it seems impossible to observe even one entire oscillation, because 
the temperature relaxation takes less time. At higher temperatures the 
situation is slightly improved. At $245^0$C the force is 10 times 
and the frequency 4.2 times larger.
\begin{flushleft}
{\bf 4. Conclusions and outlook}
\end{flushleft}

In order to discuss thermal effects, we have considered the standard setup 
for the Casimir force: two parallel, conducting square plates (side $L$) in 
a distance $a$. The plates are enclosed in a cube $L\times L\times L$ with 
one plate at face. An ensemble of free photons fufilling boundary conditions 
at the plates resp. faces was investigated. 

In section 2 we have used Quantum Statistical Mechanics to compute the 
(free energy, entropy, internal energy) per area $L^2$ and the pressure. 
We have calculated the contributions from the space between the plates 
resp. outside of them separately.\newline
In order to keep track of ultraviolet divergences, a standard regularisation 
was used. The divergences are absent both in the entropy and in the total 
pressure. They survive both in the free and internal energy in the extensive 
contribution from the external cube (cf. Table 1).\newline
Separating all extensive contributions, the remainders can be written as 
products of a factor containing some power of $a$ and a function of a 
dimensionless variable (8) (cf. (13)). These four functions are related by 
two linear equations, of which one (15) is well-known from thermodynamics. 
The other one (14) is connected with scaling properties. Both formulae 
(Table 2) and diagrams (Fig.1) for the functions are presented. The results 
allow for accurate quantitative information, also in the (relatively narrow) 
domain between the behavior at low resp. high temperatures.

In section 3 we have considered a situation with different temperatures 
between ($T$) resp. outside ($T'$) of the plates. For $T>T'$ the Casimir 
pressure is reduced by thermal effects and at some distance $a$ the total 
pressure can vanish, so that the regions inside resp. outside the plates 
are in mechanical equilibrium. If both $T$ and $T'$ are fixed (isothermal 
case), this equilibrium has turned out unstable. If only $T'$ is fixed 
and $T$ is allowed to vary with $a$ in such a way, that the entropy 
between the plates remains constant (adiabatic case), we have found also 
a region with stable equilibrium, if $T'/T(a=0)$ is small enough. The 
frequency of oscillations of one plate about stable equilibrium distances 
turned out too low for an observation (cf.(28)). Whether the (very small) 
restoring force (cf.(27)) can be measured, must be left open. 

It is known \cite{MO}, that the Casimir force depends on the geometry of 
the setup. Accurate experimental results \cite{EX2,EX3} have so far been 
obtained only for a setup consisting of a plate and a sphere. Temperature 
effects have been observed, but only the low temperature behavior was 
involved. An experiment with the original setup (as studied here), carried 
out at larger distances, would test the theory in a different domain.

\begin{flushleft}
{\bf Acknowledgments}
\end{flushleft}
We would like to thank K.Scharnhorst, G.Barton and P.Kocevar for discussions. 
Calculations were carried out using Mathematica 3.0.

\begin{appendix}
\begin{flushleft}
{\bf A. Appendix}
\end{flushleft}
Here we shall give a collection of formulae for the polylogarithm ${\cal L}$, 
which we have used in computations. Some of these are also useful in Quantum 
Statistcal Mechanics of ideal Bose gasses with other constituents than photons. 

Let $r,s$ be integers and $0\leq y\leq 1$. The polylogarithm ${\cal L}_r (y)$ 
can be defined by the series

\parbox{12cm}{
\beo
{\cal L}_r (y)=\sum_{n=1}^{\infty}\frac {y^n}{n^r}
\eeo
}\hfill
\parbox{6mm}{(A1)}

and fulfills evidently the relations

\parbox{12cm}{
\beo
{\cal L}_r(0)=0,\ \ {\cal L}_r(1)=\zeta (r),\ \ {\cal L}_r(y)\geq 0.
\eeo
}\hfill
\parbox{6mm}{(A2)}

By differentiation we obtain

\parbox{12cm}{
\beo
y\frac{d}{dy}{\cal L}_r(y)={\cal L}_{r-1}(y),
\eeo
}\hfill
\parbox{6mm}{(A3)}

\parbox{12cm}{
\beo
\int {\cal L}_r(y)\frac{dy}{y}={\cal L}_{r+1}.
\eeo
}\hfill
\parbox{6mm}{(A4)}

For $r\leq 1$ the function is elementary. We have

\parbox{12cm}{
\beo
{\cal L}_1(y)=-\ln (1-y),
\eeo
}\hfill
\parbox{6mm}{(A5)}

\parbox{12cm}{
\beo
{\cal L}_0(y)=\frac{y}{1-y}.
\eeo
}\hfill
\parbox{6mm}{(A6)}

Putting $y=\exp(-z)$, we obtain from (A3)

\parbox{12cm}{
\beo
\left( \frac{d}{dz}\right) ^s {\cal L}_r(\exp(-z))=
(-1)^s{\cal L}_{r-s}(\exp(-z)).
\eeo
}\hfill
\parbox{6mm}{(A7)}

From (A4) we obtain

\parbox{12cm}{
\beo
\int {\cal L}_r(\exp(-z))dz=-{\cal L}_{r+1}(\exp(-z)).
\eeo
}\hfill
\parbox{6mm}{(A8)}

Combining (A7) and (A8) and using repeated partial integration, we obtain

\parbox{12cm}{
\beo
\int z^s{\cal L}_r(\exp(-z))dz=-s!\sum_{n=0}^s \frac{z^{s-n}}{(s-n)!}
{\cal L}_{r+n+1}(\exp(-z)).
\eeo
}\hfill
\parbox{6mm}{(A9)}

With $r=1$ this formula can be used for the evaluation of the integrals on 
$G_\beta $ in (3),(4).

For partition functions we have to put $z=nx$ and to evaluate infinite sums 
on $n$. A useful formula reads

\parbox{12cm}{
\beo
\sum_{n=1}^{\infty }{\cal L}_r(\exp(-nx))=\sum_{n=1}^{\infty}\frac{1}{n^r}
\frac{1}{\exp (nx)-1}
\eeo
}\hfill
\parbox{6mm}{(A10)}

The second form is obtained from the first one using the definition of the 
polylogarithm and summing the resulting geometrical series. It is noted, 
that the second form converges exponentially (also for negative $r$), as 
long as $x>0$. For an approximate evaluation the series can be terminated, 
if $x$ is large enough. 

The functions (10) can be obtained starting from (A10) with $r=1$:

\parbox{12cm}{
\beo
j(x)=\sum_{n=1}^{\infty}\frac{1}{n^3N},\ \ N=\exp(nx)-1>0.
\eeo
}\hfill
\parbox{6mm}{(A11)}

Carrying out the derivatives, we obtain

\parbox{12cm}{
\beo
k(x)=(1-x\frac{d}{dx})j(x)=\sum_{n=1}^{\infty} \frac{1}{n^3}\left( 
\frac{1+nx}{N}+\frac{nx}{N^2}\right) >0,
\eeo
}\hfill
\parbox{6mm}{(A12)}

\parbox{12cm}{
\beo
h(x)=-x^2\sum_{n=1}^{\infty}\frac{1}{n}\left( \frac{1}{N}+\frac{3}{N^2}
+\frac{2}{N^3}\right) <0.
\eeo
}\hfill
\parbox{6mm}{(A13)}

These forms have been used in all numerical calculations.
\end{appendix}

\end{document}